\documentclass[conference]{IEEEtran}
\usepackage{cite}
\usepackage{amsmath,amssymb,amsfonts}
\usepackage{algorithmic}
\usepackage{graphicx}
\usepackage{textcomp}
\usepackage{xcolor}
\usepackage{tikz}
\usepackage{booktabs}
\usepackage{url}
\usepackage{multirow}
\usepackage{float}
\usepackage{hyperref}
\usepackage{listings}
\usepackage{xcolor}

\lstdefinestyle{myAsmStyle}{
    language=[x86masm]Assembler,
    basicstyle=\ttfamily\small,
    otherkeywords={>r, r>, br_if}, 
    morekeywords={print, key, dup, drop, eq, jump},
    numbers=left,
    numberstyle=\tiny\color{gray},
    stepnumber=1,
    frame=single,
    xleftmargin=2.5em,
    framexleftmargin=2em,
    breaklines=true,
    tabsize=4
}

\lstdefinestyle{myVerilogStyle}{
    language=Verilog,
    basicstyle=\ttfamily\small,
    keywordstyle=\color{blue}\bfseries,
    commentstyle=\color{green!50!black},
    stringstyle=\color{red},
    numbers=left,
    numberstyle=\tiny\color{gray},
    stepnumber=1,
    frame=single,
    xleftmargin=2.5em,      
    framexleftmargin=2em,   
    breaklines=true,
    tabsize=2
}

\usetikzlibrary{shapes,arrows,positioning,fit,calc,automata,shadows}
\tikzset{
    block/.style = {draw, rectangle, minimum height=2.5em, minimum width=4em, align=center, fill=blue!5, drop shadow},
    memory/.style = {draw, rectangle, minimum height=4em, minimum width=3em, align=center, fill=yellow!10, drop shadow},
    line/.style = {->, thick, >=stealth'},
    state_node/.style = {circle, draw, minimum size=1.2cm, fill=green!5, font=\footnotesize, align=center, drop shadow}
}

\begin{document}

\title{A WASM-Subset Stack Architecture for Low-cost FPGAs using Open-Source EDA Flows}

\author{\IEEEauthorblockN{Aradhya Chakrabarti}
\IEEEauthorblockA{\textit{School of Computer Engineering} \\
\textit{KIIT Deemed to be University}\\
Bhubaneswar, India \\
2205880@kiit.ac.in}}

\maketitle

\begin{abstract}
Soft-core processors on resource-constrained FPGAs often suffer from low code density and reliance on proprietary toolchains. This paper details the design, implementation, and evaluation of a 32-bit dual-stack microprocessor architecture optimized for low-cost, resource-constrained Field-Programmable Gate Arrays (FPGAs). Implemented on the Gowin GW1NR-9 (Tang Nano 9K), the processor utilizes an instruction set architecture (ISA) inspired from a subset of the WebAssembly (WASM) specification to achieve high code density. Unlike traditional soft-cores that often rely on proprietary vendor toolchains and opaque IP blocks, this design is synthesized and routed utilizing an open-source flow, providing transparency and portability. The architecture features a dual-stack model (Data and Return), executing directly from SPI Flash via an Execute-in-Place (XIP) mechanism to conserve scarce Block RAM on the intended target device. An analysis of the trade-offs involved in stack depth parametrization is presented, demonstrating that an 8-entry distributed RAM implementation provides a balance between logic resource utilization ($\sim$80\%) and routing congestion. Furthermore, timing hazards in single-cycle stack operations are identified and resolved through a refined Finite State Machine (FSM) design. The system achieves a stable operating frequency of 27 MHz, limited by Flash latency, and successfully executes simple applications including a single and multi-digit infix calculator.
\end{abstract}

\begin{IEEEkeywords}
FPGA, Stack Machine, WebAssembly, Open Source EDA, Gowin, Soft-core, Embedded Systems.
\end{IEEEkeywords}

\section{Introduction}

The proliferation of Field-Programmable Gate Arrays (FPGAs) in embedded systems has created a demand for soft-core processors that can be instantiated alongside custom logic. While commercial cores like Nios II \cite{ball2007designing} and MicroBlaze offer high performance, they often consume significant resources and rely on complex register-based architectures. Furthermore, their dependency on proprietary toolchains limits their utility in open hardware research and education.

Stack machines, popularized by the Forth programming language and the Java Virtual Machine (JVM), offer a compelling alternative. By eliminating explicit operand addressing, stack architectures achieve superior code density and simpler control logic \cite{koopman1990modern}. The recent adoption of WebAssembly (WASM) as a standard bytecode format provides a modern, well-specified stack ISA that is ideal for hardware implementation.

This work focuses on the challenge of implementing a competent 32-bit processor on the Gowin GW1NR-9, a low-cost FPGA with only 8,640 logic elements (LUTs). This constraint forces rethinking of standard design patterns (like large register files or deep pipelines) and necessitates a more bare-metal approach to architecture.

Specific contributions include:
\begin{enumerate}
    \item \textbf{Microarchitecture}: A complete Verilog HDL Register Transfer Level (RTL) implementation \cite{github_repo} of a WASM-like subset soft-core that fits within 2,000 logic elements (excluding interconnect), featuring a novel 12-state FSM to handle variable-length instruction fetching.
    \item \textbf{Timing Analysis}: A detailed analysis of race conditions which occur in single-cycle stack operations and a simple solution using a separate wait-state mechanism.
    \item \textbf{Resource Optimization}: An experimental analysis of stack depth implementation strategies (Distributed RAM vs. Block RAM) and their impact on routing congestion for this design.
    \item \textbf{Toolchain}: A custom two-pass assembler enabling the translation of assembly language code (based on the implemented ISA) into binary images for SPI Flash.
\end{enumerate}

\section{Related Work}

\subsection{Stack Architecture Fundamentals}
The theoretical foundations of modern stack computers were established by Koopman \cite{koopman1990modern}, who argued that dual-stack architectures which separate data calculations from control flow return addresses, offer superior interrupt latency and context switching performance compared to Register-based machines. The presented design utilizes this dual-stack model, implementing separate Data and Return stacks to allow for Forth-style control flow and efficient subroutine linkage.

\subsection{Advanced Stack Machines}
Schoeberl's \cite{schoeberl2005design} advanced stack machine design with the Java Optimized Processor (JOP), introducing a "stack cache" to bridge the gap between on-chip memory and the ALU. While Schoeberl's work focuses on the complex Java Virtual Machine (JVM), the presented work targets a simpler, WASM-like model. Unlike the PicoJava \cite{mcghan1998picojava} implementation, which included hardware support for garbage collection and object management (occupying significant silicon area), this design focuses on simplicity suitable for resource-constrained FPGAs, with deterministic timing.

\subsection{FPGA Soft-Cores}
Ball \cite{ball2007designing} provides a detailed analysis of Altera's Nios II, noting that soft-core efficiency is defined by the ratio of performance to logic element (LE) cost. He highlights that while FPGA RAM blocks are abundant, logic elements are precious. However, utilizing Block RAM (BRAM) for small register files can be wasteful of memory bandwidth. The presented design contrasts with this by using distributed LUT-RAM for stack storage, a technique viable only for shallow stacks, preserving BRAM for main data memory.

Educational processors like the TINYCPU \cite{nakano2008processor} and simple 8-bit processors described by Upadhyaya \cite{upadhyayadesign} and Ayeh \cite{ayeh2008fpga} demonstrate the value of building processors from scratch. However, these designs often utilize custom, non-standard ISAs with limited software ecosystem potential. By adopting a WASM-like ISA, the presented design intends to align with a global standard, opening the door for future compiler interoperability.

\subsection{The Open Source Ecosystem}
The reverse engineering of the Gowin bitstream format by De Vos et al. \cite{de2020complete} allowed for the usage of an open source and transparent toolchain for the implementation of the presented design. This effort allowed the integration of Gowin devices into the Yosys/nextpnr ecosystem. The architecture for the presented design is written in standard Verilog-2001, ensuring portability across both proprietary and open-source flows. Additionally, the Tang Nano 9K platform documentation by Lushay Labs \cite{lushaylabs} provided essential guidance for board-specific constraints.

\section{Architecture Comparison: Stack vs. Register Machines}

A fundamental decision in any processor design is the choice of operand storage. Most modern general-purpose processors (like x86, ARM, RISC-V) utilize a register-based architecture, while this work implements a stack-based architecture. The differences are significant, particularly in the context of resource-constrained FPGAs.

\subsection{Operand Addressing and Code Density}
In a register machine, arithmetic instructions must explicitly specify source and destination registers (e.g., \texttt{ADD R1, R2, R3}). This necessitates larger instruction words (typically 32 bits) to encode these register indices. In contrast, stack machines use zero-address instructions (e.g., \texttt{ADD}). Operands are implicitly consumed from the top of the stack, and the result is pushed back. This implicit addressing allows for extremely compact instruction encoding, often just a single byte \cite{koopman1990modern}. For our FPGA implementation, where program memory is fetched from a slow serial Flash, this high code density directly translates to improved effective throughput.

\subsection{Hardware Complexity}
Register machines require a multi-ported register file to sustain instruction throughput (typically 2 read ports and 1 write port per cycle). Implementing such multi-ported memories on an FPGA using logic elements (Distributed RAM) is expensive in terms of area and routing resources \cite{ball2007designing}. A stack machine, however, only requires access to the top two elements of the stack. By caching just these top elements in registers (or using a shallow distributed RAM as done in this work), the hardware complexity is significantly reduced. This aligns with the findings of Schoeberl \cite{schoeberl2005design}, who noted that a stack cache simplifies the pipeline by eliminating the need for complex data forwarding logic common in RISC pipelines.

\subsection{Context Switching and Interrupts}
One of the advantages of stack architectures is their lightweight context switching \cite{koopman1990modern}. In a register machine, an interrupt requires saving the entire register set to memory, which consumes cycles and memory bandwidth. In a pure stack machine, the stack is already in memory (or can be treated as such), so context switching only requires saving a few pointers (Stack Pointer, Program Counter). While this implementation uses a small on-chip stack, there is no large register file state to preserve, making interrupt entry and exit faster and simpler to implement in hardware.

\section{Instruction Set Architecture}

The ISA is a 32-bit adaptation of WebAssembly \cite{wasm_spec}, designed to be executed directly in hardware. It enforces a stack-based execution model where all arithmetic and logical operations consume operands from the top of the Data Stack and push results back.

\subsection{Instruction Encoding}
To maximize code density on the SPI Flash (which has high latency), a variable-length instruction encoding is employed:

\begin{itemize}
    \item \textbf{Single-Byte Instructions (Opcode Only):} The majority of instructions (e.g., \texttt{ADD}, \texttt{SUB}, \texttt{DUP}, \texttt{DROP}) are encoded as a single byte. This allows the CPU to execute tight loops with minimal instruction fetch overhead.
    \item \textbf{Multi-Byte Instructions (Opcode + Immediate):} Instructions requiring constants or branch targets (e.g., \texttt{PUSH}, \texttt{BR\_IF}, \texttt{CALL}) consist of a 1-byte opcode followed by a 32-bit Little-Endian immediate value. This totals 5 bytes per instruction.
\end{itemize}

\subsection{Complete Operation Table}
The processor implements 40 instructions, mapped to the standard WASM opcodes where applicable. Table \ref{tab:isa} details the core instruction set.

\begin{table}[H]
\caption{Core Instruction Set Architecture}
\begin{center}
\begin{tabular}{l c l l}
\toprule
\textbf{Mnemonic} & \textbf{Opcode} & \textbf{Stack Effect} & \textbf{Description} \\
\midrule
\multicolumn{4}{l}{\textit{Stack Manipulation}} \\
\texttt{PUSH imm} & 0x01 & $( - n )$ & Push 32-bit constant \\
\texttt{DROP} & 0x05 & $( n - )$ & Discard top value \\
\texttt{DUP} & 0x12 & $( n - n \ n )$ & Duplicate top value \\
\texttt{SWAP} & 0x13 & $( a \ b - b \ a )$ & Swap top two values \\
\texttt{OVER} & 0x14 & $( a \ b - a \ b \ a )$ & Copy 2nd item to top \\
\midrule
\multicolumn{4}{l}{\textit{Arithmetic \& Logic}} \\
\texttt{ADD} & 0x02 & $( a \ b - a+b )$ & 32-bit addition \\
\texttt{SUB} & 0x03 & $( a \ b - a-b )$ & 32-bit subtraction \\
\texttt{MUL} & 0x04 & $( a \ b - a*b )$ & 32-bit multiplication \\
\texttt{AND} & 0x16 & $( a \ b - a\&b )$ & Bitwise AND \\
\texttt{OR} & 0x17 & $( a \ b - a|b )$ & Bitwise OR \\
\texttt{NOT} & 0x19 & $( n - \sim n )$ & Bitwise NOT \\
\midrule
\multicolumn{4}{l}{\textit{Comparison (Returns 1 for True, 0 for False)}} \\
\texttt{EQ} & 0x09 & $( a \ b - c )$ & Equality check \\
\texttt{LT\_S} & 0x0A & $( a \ b - c )$ & Signed Less Than \\
\texttt{GT\_S} & 0x0B & $( a \ b - c )$ & Signed Greater Than \\
\texttt{EQZ} & 0x35 & $( n - c )$ & True if zero \\
\midrule
\multicolumn{4}{l}{\textit{Control Flow}} \\
\texttt{BR\_IF} & 0x0E & $( c - )$ & Branch if top != 0 \\
\texttt{JUMP} & 0x0F & $( - )$ & Unconditional Jump \\
\texttt{CALL} & 0x10 & $( - )$ & Call subroutine \\
\texttt{RET} & 0x11 & $( - )$ & Return from sub \\
\midrule
\multicolumn{4}{l}{\textit{Memory \& I/O}} \\
\texttt{LOAD} & 0x1D & $( addr - val )$ & Load word from RAM \\
\texttt{STORE} & 0x1E & $( val \ addr - )$ & Store word to RAM \\
\texttt{PRINT} & 0x08 & $( n - )$ & UART TX (low byte) \\
\texttt{KEY} & 0x1F & $( - char )$ & Blocking UART RX \\
\bottomrule
\end{tabular}
\label{tab:isa}
\end{center}
\end{table}

\section{Microarchitecture Implementation}

\subsection{System Overview}
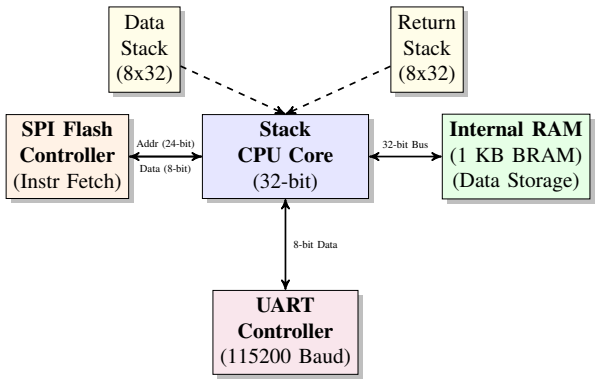
\begin{figure}[H]
\centering
\resizebox{0.9\columnwidth}{!}{
\begin{tikzpicture}[auto, node distance=1.5cm]
    \node [block, text width=2.5cm, fill=blue!10] (cpu) {\textbf{Stack CPU Core} \\ (32-bit)};
    \node [block, left=1.2cm of cpu, fill=orange!10] (flash) {\textbf{SPI Flash} \\ \textbf{Controller} \\ (Instr Fetch)};
    \node [block, right=1.2cm of cpu, fill=green!10] (ram) {\textbf{Internal RAM} \\ (1 KB BRAM) \\ (Data Storage)};
    \node [block, below=1.5cm of cpu, fill=purple!10] (uart) {\textbf{UART} \\ \textbf{Controller} \\ (115200 Baud)};
    
    \node [memory, above left=0.5cm of cpu, anchor=south east] (dstack) {Data \\ Stack \\ (8x32)};
    \node [memory, above right=0.5cm of cpu, anchor=south west] (rstack) {Return \\ Stack \\ (8x32)};
    
    \draw [line] (cpu) -- node[above, font=\tiny] {Addr (24-bit)} (flash);
    \draw [line] (flash) -- node[below, font=\tiny] {Data (8-bit)} (cpu);
    
    \draw [line, <->] (cpu) -- node[above, font=\tiny] {32-bit Bus} (ram);
    \draw [line, <->] (cpu) -- node[right, font=\tiny] {8-bit Data} (uart);
    
    \draw [line, dashed] (dstack) -- (cpu.north);
    \draw [line, dashed] (rstack) -- (cpu.north);
    
\end{tikzpicture}
}
\caption{System Architecture utilizing the Tang Nano 9K resources. The CPU implements a Harvard-like split with instructions in Flash and data in SRAM.}
\label{fig:arch}
\end{figure}

The system architecture (Figure \ref{fig:arch}) is centered around the Stack CPU core, which acts as the bus master. It interfaces with three distinct subsystems:
\begin{enumerate}
    \item \textbf{SPI Flash Controller}: Fetches instructions. Due to the high latency of SPI (serial) access, instruction fetching is the primary bottleneck.
    \item \textbf{Internal SRAM}: A 1KB Block RAM used for data storage (variables, buffers). It allows single-cycle access.
    \item \textbf{UART Controller}: Provides standard I/O capability at 115200 baud.
\end{enumerate}

\subsection{Finite State Machine (FSM)}
The control unit is implemented as a 12-state FSM. The FSM describes variable-length instruction fetching and the management of multi-cycle I/O and memory operations.

The FSM states are:
\begin{itemize}
    \item \textbf{FETCH Sequence (3 states):} Handles the SPI protocol to read 1 byte from Flash. Includes \texttt{FETCH\_WAIT\_LOW} and \texttt{FETCH\_WAIT\_HIGH} to accommodate Flash latency.
    \item \textbf{DECODE:} Decodes the opcode. If the instruction requires an immediate value (e.g., \texttt{PUSH}), transitions to the \texttt{FETCH\_IMM} loop. Otherwise, transitions to \texttt{EXECUTE}.
    \item \textbf{FETCH\_IMM Sequence:} A loop that runs 4 times to assemble a 32-bit immediate value byte-by-byte from Flash.
    \item \textbf{EXECUTE:} The primary cycle where ALU operations, stack pointer updates, and memory requests occur.
    \item \textbf{ALU\_WAIT:} A delay state for comparison operations (see Section \ref{sec:race}).
    \item \textbf{I/O Waits:} \texttt{UART\_WAIT} for transmission and \texttt{KEY\_WAIT} for blocking reception.
\end{itemize}

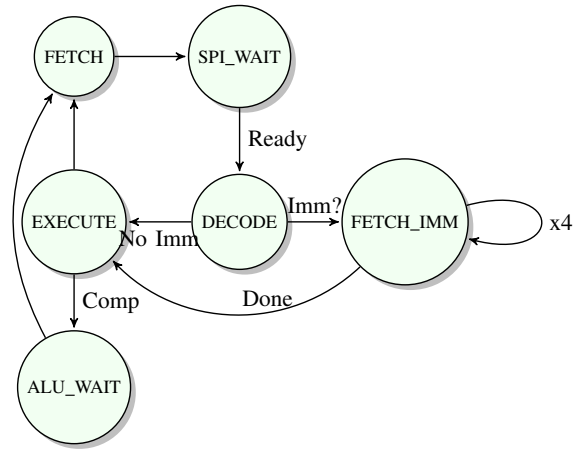
\begin{figure}[H]
\centering
\resizebox{0.9\columnwidth}{!}{
\begin{tikzpicture}[->,>=stealth',shorten >=1pt,auto,node distance=2.5cm, semithick]
  \node[state_node] (FETCH) {FETCH};
  \node[state_node] (WAIT) [right of=FETCH] {SPI\_WAIT};
  \node[state_node] (DECODE) [below of=WAIT] {DECODE};
  \node[state_node] (EXEC) [left of=DECODE] {EXECUTE};
  \node[state_node] (ALU_W) [below of=EXEC] {ALU\_WAIT};
  \node[state_node] (IMM) [right of=DECODE] {FETCH\_IMM};

  \path (FETCH) edge node {} (WAIT)
        (WAIT) edge node {Ready} (DECODE)
        (DECODE) edge node {Imm?} (IMM)
        (DECODE) edge node {No Imm} (EXEC)
        (IMM) edge [loop right] node {x4} (IMM)
        (IMM) edge [bend left=45] node[above right] {Done} (EXEC) 
        (EXEC) edge node {} (FETCH)
        (EXEC) edge node {Comp} (ALU_W)
        (ALU_W) edge [bend left] node {} (FETCH);
\end{tikzpicture}
}
\caption{Simplified FSM describing the fetch-decode-execute loop.}
\label{fig:fsm}
\end{figure}

\subsection{Data Path \& Stack Implementation}
A defining architectural decision during the implementation of the presented design was the implementation of the stacks. \textbf{Distributed RAM} (LUT-based memory) was chosen over over Block RAM.

\begin{itemize}
    \item \textbf{Data Stack:} 8 entries deep, 32-bits wide.
    \item \textbf{Return Stack:} 8 entries deep, 32-bits wide.
\end{itemize}

While modern FPGAs have Block RAM, utilizing a 9Kbit block for a 512-bit stack is inefficient. Furthermore, Distributed RAM allows for asynchronous read access (effectively zero-latency in the RTL model) which simplifies the control logic. The stack pointers are 3-bit registers that wrap around automatically, providing a circular buffer effect.

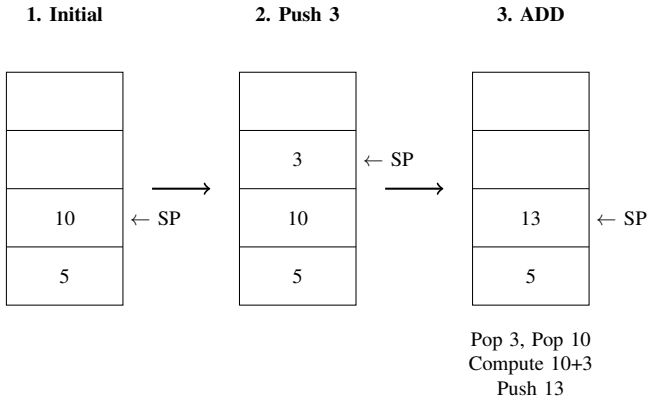
\begin{figure}[H]
\centering
\resizebox{\columnwidth}{!}{
\begin{tikzpicture}[scale=0.8, transform shape]
    \node at (1, 5) {\textbf{1. Initial}};
    \draw (0,0) rectangle (2,4);
    \foreach \y in {1,2,3} \draw (0,\y) -- (2,\y);
    \node at (1, 0.5) {5};
    \node at (1, 1.5) {10};
    \node[anchor=west] at (2, 1.5) {$\leftarrow$ SP};

    \node at (5, 5) {\textbf{2. Push 3}};
    \draw (4,0) rectangle (6,4);
    \foreach \y in {1,2,3} \draw (4,\y) -- (6,\y);
    \node at (5, 0.5) {5};
    \node at (5, 1.5) {10};
    \node at (5, 2.5) {3};
    \node[anchor=west] at (6, 2.5) {$\leftarrow$ SP};
    \draw[->, thick] (2.5, 2) -- (3.5, 2);

    \node at (9, 5) {\textbf{3. ADD}};
    \draw (8,0) rectangle (10,4);
    \foreach \y in {1,2,3} \draw (8,\y) -- (10,\y);
    \node at (9, 0.5) {5};
    \node at (9, 1.5) {13};
    \node[anchor=west] at (10, 1.5) {$\leftarrow$ SP};
    \draw[->, thick] (6.5, 2) -- (7.5, 2);
    
    \node[text width=4cm, align=center] at (9, -1) {Pop 3, Pop 10\\Compute 10+3\\Push 13};

\end{tikzpicture}
}
\caption{An Exemplar Visualization of the Data Stack during a \texttt{PUSH 3} followed by an \texttt{ADD} operation. The Stack Pointer (SP) moves up and down as data is pushed and popped.}
\label{fig:stack_viz}
\end{figure}

\subsection{Handling Race Conditions} \label{sec:race}
A significant challenge in single-cycle stack machines is the \texttt{Read-Modify-Write} hazard on the stack memory. Consider the comparison instruction \texttt{EQ} (Equal), which pops two values and pushes a boolean result.
\\\\
\textbf{Initial Implementation:}
\begin{lstlisting}[style=myVerilogStyle, caption={Initial implementation illustrating the race condition issue.}, label={lst:incorrect_verilog}]
// Inside EXECUTE state
// Compare top two items
temp_val = (stack[sp-1] == stack[sp-2]); 
sp <= sp - 1;          // Decrement stack pointer
stack[sp] <= temp_val; // Write result to NEW top
\end{lstlisting}

In hardware, \texttt{sp} does not update instantly. If \texttt{stack[sp]} uses the \texttt{sp} signal as its address, it might write to the \textit{old} address before \texttt{sp} decrements, or worse, the comparator logic might see the stack outputs changing as \texttt{sp} changes, leading to a bug. This shows the limitations of treating hardware description like sequential software code.
\\\\
\textbf{Optimized State Machine (ALU\_WAIT):}
\\\\
In order to address this, a dedicated state, \texttt{ALU\_WAIT} was introduced.
\begin{itemize}
    \item \textbf{Cycle 1 (EXECUTE):} Perform the comparison combinatorially and latch the result into a temporary register (\texttt{temp\_alu}). Decrement the stack pointer register.
    \item \textbf{Cycle 2 (ALU\_WAIT):} The stack pointer is now stable at its new value ($sp-1$). Write the latched \texttt{temp\_alu} result to \texttt{stack[sp]}.
\end{itemize}

This ensures data integrity at the cost of one extra clock cycle for comparison operations.

\section{Assembler \& Toolchain}

To program the core, a custom assembler was developed in JavaScript (Node.js). The assembler translates between the high-level WASM-like assembly syntax and the binary required by the FPGA.

\subsection{Two-Pass Architecture}
The assembler uses a two-pass approach to handle \textbf{forward label references}, a common feature in structured programming. An example source code in the implemented format is provided in Appendix \ref{sample_code}.

\begin{enumerate}
    \item \textbf{Pass 1 (Label Collection):} The assembler scans the source file. It calculates the byte offset of every instruction but generates no code. When a label definition (e.g., \texttt{:loop}) is encountered, its calculated address is stored in a symbol table.
    \item \textbf{Pass 2 (Code Generation):} The assembler re-scans the file. It looks up label targets in the symbol table. For instructions like \texttt{br\_if :label}, it calculates the numerical immediate value and emits the binary sequence.
\end{enumerate}

\subsection{Immediate Encoding}
The assembler handles the Little-Endian conversion automatically. For the instruction \texttt{push 0x12345678}, the assembler generates the byte sequence:
\texttt{01 78 56 34 12}
Where \texttt{01} is the opcode for PUSH, and the following bytes are the immediate value LSB first.

\section{Implementation Results}

\subsection{Resource Utilization}
The design was synthesized for the Gowin GW1NR-9 FPGA using the \href{https://github.com/YosysHQ/oss-cad-suite-build}{OSS-CAD Suite} (Yosys/Nextpnr/OpenFPGALoader).

\begin{table}[H]
\caption{Resource Utilization on Tang Nano 9K}
\begin{center}
\begin{tabular}{l c c c}
\toprule
\textbf{Resource} & \textbf{Used} & \textbf{Total Available} & \textbf{Percentage} \\
\midrule
Logic (LUT4) & 7,350 & 8,640 & 80\% \\
Registers (FF) & 2,100 & 6,480 & 32\% \\
Block RAM & 2 & 26 & 8\% \\
\bottomrule
\end{tabular}
\label{tab:resources}
\end{center}
\end{table}

The high LUT utilization (80\%) is caused mainly by the multiplexers required for the 32-bit datapath and the distributed RAM implementation of the stacks. The low Block RAM usage indicates significant headroom for increasing the data memory size (up to ~26KB).

\subsection{Stack Depth Trade-off Analysis}
A sensitivity analysis was performed on the stack depth parameter to find the optimal configuration for this specific FPGA.

\begin{enumerate}
    \item \textbf{4-Entry Stack:} Consumed ~65\% logic. However, complex programs like the calculator caused stack overflows due to insufficient depth for intermediate results.
    \item \textbf{8-Entry Stack:} Consumed ~80\% logic. This provided enough depth for all tested applications while remaining routable.
    \item \textbf{16-Entry Stack:} Consumed ~99\% logic. The routing tools failed to meet timing closure due to congestion.
\end{enumerate}

Thus, 8 entries (3-bit pointer) was determined to be the local optimum for the GW1NR-9 architecture.

\subsection{Timing and Performance}
\begin{itemize}
    \item \textbf{Max Frequency:} 27 MHz. The design is constrained by the SPI Flash read latency ($\sim$200ns). The internal logic can run significantly faster (estimated \(F_{max}\) \(>\) 40 MHz), but the fetch bottleneck is significant.
    \item \textbf{Throughput:} 4-6 MIPS. Simple instructions take 5 cycles (fetch + decode + execute). Instructions with immediates take 17 cycles due to the 4-byte fetch sequence.
\end{itemize}

\subsection{I/O Timing Analysis}

\begin{figure}[H]
\centering
\resizebox{\columnwidth}{!}{
\begin{tikzpicture}[xscale=0.5, yscale=0.8]
    \draw[help lines, lightgray] (0,0) grid (22, 2);
    
    \node at (-1, 1.5) {TX Line};
    \node at (11, -0.5) {Time ($\mu s$)};
    
    \draw[thick] (0,1.5) -- (2,1.5); 
    \node[above] at (1, 1.5) {IDLE};
    
    \draw[thick] (2,1.5) -- (2,0.5) -- (4,0.5);
    \node[above] at (3, 0.5) {START};
    \node[below] at (3, 0.5) {0};
    
    \draw[thick] (4,0.5) -- (4,1.5) -- (6,1.5); \node[below] at (5, 1.5) {1};
    \draw[thick] (6,1.5) -- (6,0.5) -- (8,0.5); \node[below] at (7, 0.5) {0};
    \draw[thick] (8,0.5) -- (10,0.5); \node[below] at (9, 0.5) {0};
    \draw[thick] (10,0.5) -- (12,0.5); \node[below] at (11, 0.5) {0};
    \draw[thick] (12,0.5) -- (14,0.5); \node[below] at (13, 0.5) {0};
    \draw[thick] (14,0.5) -- (16,0.5); \node[below] at (15, 0.5) {0};
    \draw[thick] (16,0.5) -- (16,1.5) -- (18,1.5); \node[below] at (17, 1.5) {1};
    \draw[thick] (18,1.5) -- (18,0.5) -- (20,0.5); \node[below] at (19, 0.5) {0};
    
    \draw[thick] (20,0.5) -- (20,1.5) -- (22,1.5);
    \node[above] at (21, 1.5) {STOP};
    
    \draw[<->] (2, 2) -- (4, 2);
    \node[above] at (3, 2) {8.68$\mu s$};

\end{tikzpicture}
}
\caption{UART Transmission Timing Diagram for character 'A' (0x41) at 115200 baud. The signal is active low for the Start bit and logic 0 data bits.}
\label{fig:uart_timing}
\end{figure}
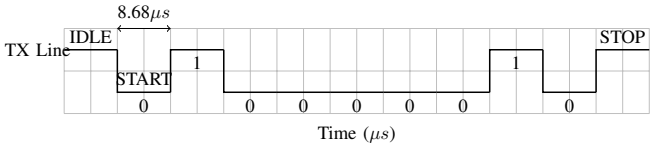

Asynchronous protocols like UART require precise timing. The CPU operates at 27 MHz, and the UART module is designed to transmit at 115200 baud. Figure \ref{fig:uart_timing} illustrates the signal timing for transmitting the character 'A' (0x41).

\section{Case Study: Calculator}

To evaluate the functional correctness and control-flow capability of the processor, a full Infix calculator was implemented in assembly. This application exercises arithmetic, stack manipulation, UART I/O, and control flow. The MIT-licensed source code is available in the source code repository for this implementation \cite{github_repo}.

\subsection{Functionality}
The calculator accepts inputs like \texttt{1+2} via UART and prints \texttt{3}. It supports multi-digit parsing (e.g., "123") which requires a loop:
$$ \text{value} = (\text{value} \times 10) + (\text{char} - \text{'0'}) $$
This loop shows the stack's ability to hold the accumulator, the address pointer, and the incoming character simultaneously.

\subsection{Software Division}
Since the core lacks a hardware divider (to save space), division is implemented in software using repeated subtraction. The \texttt{lt\_s} (signed less than) and \texttt{br\_if} instructions are critical here. The successful execution of this routine validates the correctness of the comparison logic and the \texttt{ALU\_WAIT} fix.

\begin{figure}
    \centering
    \includegraphics[width=0.5\linewidth]{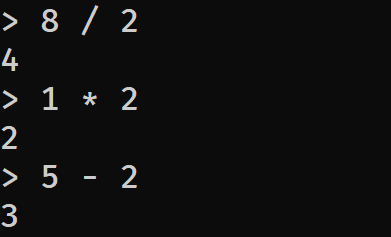}
    \caption{Sample single-digit calculator execution output}
    \label{lst:calc_session}
\end{figure}

\section{Conclusion}

This work demonstrates that modern, open-source-friendly EDA flows are capable of supporting complex soft-core designs even on entry-level FPGAs. By adhering to a stristack-based architecture, we achieved a high degree of functionality, including a 32-bit datapath and full I/O capability was achieved with footprint of less than 8,000 LUTs.

This implementation validates Koopman's and Schoeberl's theories on stack machine efficiency in a modern context. The dual-stack model proved effective for managing control flow without the register-saving overhead of RISC architectures. Furthermore, the identification of the ALU race condition shows the subtle challenges of translating abstract stack models into physical RTL.

Future work will focus on implementing an instruction cache (using the abundant spare Block RAM) to decouple the CPU frequency from the SPI Flash latency, potentially doubling the throughput. Furthermore, there are plans to implement Forth as well.

\section*{Acknowledgment}
The author acknowledges the comprehensive documentation provided by Lushay Labs \cite{lushaylabs} on the Tang Nano series, which was instrumental in the board bring-up. They also acknowledge the WebAssembly Community Group for the ISA specification \cite{wasm_spec}. The complete source code and design files are available on the author's GitHub repository \cite{github_repo}.

\appendices

\section{Sample Compiled Binary}
\begin{lstlisting}[
    caption={Hex dump of the single-digit calculator application.},
    label={lst:calc_hex},
    basicstyle=\ttfamily\scriptsize, 
    columns=flexible,
    breaklines=true,
    breakatwhitespace=true,
    frame=single,
    xleftmargin=2.5em,      % <--- Adds padding from the left page margin
    framexleftmargin=2em    % <--- Extends the frame left to include the numbers
]
01 3e 00 00 00 08 01 20 00 00 00 08 1f 12 08 01
30 00 00 00 03 1f 12 08 1f 12 08 01 30 00 00 00
03 01 0d 00 00 00 08 01 0a 00 00 00 08 13 12 01
2b 00 00 00 09 0e 5a 00 00 00 12 01 2d 00 00 00
09 0e 74 00 00 00 12 01 2a 00 00 00 09 0e 8e 00
00 00 05 05 05 0f 00 00 00 00 05 02 01 30 00 00
00 02 08 01 0d 00 00 00 08 01 0a 00 00 00 08 0f
00 00 00 00 05 03 01 30 00 00 00 02 08 01 0d 00
00 00 08 01 0a 00 00 00 08 0f 00 00 00 00 05 04
01 30 00 00 00 02 08 01 0d 00 00 00 08 01 0a 00
00 00 08 0f 00 00 00 00
\end{lstlisting}

\section{Sample Application Code}\label{sample_code}
The following assembly listing demonstrates the implementation of the core logic for a simple single-digit infix calculator (Add, Subtract, Multiply). It showcases the use of I/O instructions, stack manipulation, and conditional branching.

\begin{lstlisting}[style=myAsmStyle, caption={Single-Digit Calculator Assembly Code.}, label={lst:calc_asm}]
:main
    push 62 ; '>'
    print
    push 32
    print

    key     ; digit1_char
    dup
    print
    push 48
    sub     ; digit1
    
    key     ; digit1 op_char
    dup
    print
    
    key     ; digit1 op digit2_char
    dup
    print
    push 48
    sub     ; digit1 op digit2
    
    push 13
    print
    push 10
    print
    
    ; Stack: digit1 op digit2
    ; Rearrange to: digit1 digit2 op
    swap    ; digit1 digit2 op
    
    ; Check if op is '+'
    dup     ; digit1 digit2 op op
    push 43
    eq      ; digit1 digit2 op (op==43)
    br_if :add_op
    
    ; Check if op is '-'
    dup
    push 45
    eq
    br_if :sub_op
    
    ; Check if op is '*'
    dup
    push 42
    eq
    br_if :mul_op
    
    ; Unknown - drop everything
    drop
    drop
    drop
    jump :main

:add_op
    drop    ; digit1 digit2
    add
    push 48 ; Convert to ASCII
    add
    print
    push 13 ; CR
    print
    push 10 ; LF
    print
    jump :main ; loop forever

:sub_op
    drop
    sub
    push 48
    add
    print
    push 13
    print
    push 10
    print
    jump :main

:mul_op
    drop
    mul
    push 48
    add
    print
    push 13
    print
    push 10
    print
    jump :main
\end{lstlisting}

\bibliographystyle{IEEEtran}
\bibliography{references}

\end{document}